\documentstyle[12pt]{article}
\begin{document}
\begin{center}
{\large \bf 
HERD BEHAVIOR OF RETURNS IN THE FUTURES EXCHANGE MARKET 
}\\

\vspace*{.5in}

\normalsize 
Kyungsik Kim$^{*}$, Seong-Min Yoon$^{a}$ and Yup Kim$^{b}$ \\

\vspace*{.2in}

{\em 
$^{*}$$Department$ $of$ $Physics$, $Pukyong$ $National$ $University$,\\
$Pusan$ $608$-$737$, $Korea$\\
$^{a}$$Division$ $of$ $Economics$, $Pukyong$ $National$ $University$,\\
$Pusan$ $608$-$737$, $Korea$ \\
$^{b}$$Department$ $of$ $Physics$, $Kyung$ $Hee$ $University$,\\
$Seoul$ $130$-$701$, $Korea$ \\
}

\hfill\\
\end{center} 
%
 
%
%
%
\baselineskip 24pt
\begin{center}
{\bf Abstract}
\end{center} 

\noindent
The herd behavior of returns is investigated in Korean futures exchange market. 
It is obtained that the probability distribution of returns for three types of herding parameter
scales as a power law $R^{-\beta}$ with the exponents $ \beta=3.6$(KTB$203$) and $2.9$(KTB$209$)
in two kinds of Korean treasury bond.
For our case since the active state of transaction exists to decrease
lesser than the herding parameter $h=2.33$, the crash regime appears to increase
in the probability with high returns values.
Especially, we find that it shows a crossover toward a Gaussian probability function near the time step 
$\Delta t=360$ from the distribution of normalized returns. 
Our result will be also compared with other well-known results.

\vskip 35mm
\noindent
$Keywords$: Herd behavior; Returns; Bond future prices  \\                                                        
\vskip 5mm
\noindent
$^{*}$Corresponding author. Tel.: +82-51-620-6354; Fax: +82-51-611-6357.\\
$^{*}$$E-mail$ $address$: kskim@pknu.ac.kr.

\newpage

\noindent
{\bf 1. INTRODUCTION}
\hfill\\

In financial markets, there has recently been attracted the considerable interests for the microscopic models
in natural and social sciences.$^{1-3}$
It is in particular well-known that the major models of interest in self-organized phenomena
are the herding multiagent model$^{4,5}$ and the related percolation models,$^{6,7}$ the democracy and dictatorship model,$^{8}$
self-organized dynamical model,$^{9}$ the cut and paste model, the fragmentation and 
coagulation model.$^{10}$
One of challenging microscopic models is the herding model$^{11,12}$ that means some degree of coordination and crowd effect
between a group of agents sharing the same information or the same rumor and making a common decision.  
It has recently been introduced that the probability distribution of returns 
scales a power law and that it exists the financial crashes 
in the probability with low herding parameters and high return values.$^{5}$
Moreover, the distribution of normalized returns has
the form of the fat-tailed distributions$^{13}$ of price return, and 
a crossover toward the Gaussian distribution can be shown in financial markets. 

The theoretical and numerical arguments for the volume of bond futures traded at Korean futures exchange market
were presented in the previous work.$^{14}$
We mainly considered the number of transactions for two different delivery dates and found
the decay functions for survival probabilities$^{15,16}$ in our bond futures model.
We also have studied the tick dynamical behavior of the bond futures price using
the range over standard deviation or the R/S analysis in Korean Futures Exchange market.$^{17}$ 
The Norwegian and US stock markets presented in recent work$^{18}$ have been led to the
notable persistence caused by long-memory in the time series.
The multifractal Hurst exponents and the height-height correlation 
function have mainly been discussed numerically
with long-run memory effects. It is particularly shown that the form of the probability distribution of the prices
leads to the Lorentz distribution rather than the Gaussian distribution.$^{17}$ 

The purpose of this paper is to study the dynamical herding behavior of the bond futures price
for two kinds of Korean treasury bond in Korean futures exchange market.
In this paper, we only consider two different delivery dates: October KTB$203$ and April KTB$209$. 
The tick data for KTB$203$ were taken from October $2002$ to March $2002$
while we used the tick data of March KTB$209$ transacted for six months from April $2002$.
In section $2$ we mainly find the financial crashes and the distribution of normalized returns
from the distribution of returns. We end with some results and conclusions in the final section.\\  

\noindent
{\bf 2. FINANCIAL CRASHES AND SIMULATIONS}
\hfill\\

In our model, we introduce bond futures prices for two sets of data(KTB$203$ and KTB$209$)
in Korean futures exchange market. 
In Fig.$1$, we show the time series of bond futures price $P(t)$ that is found
in the case of bond futures of KTB$203$ traded for six months at Korean futures exchange market.
We can also obtain numerically the price return $R(t_i ) = $ ln$[P(t_i )/P(t_i -1)]$,
as plotted in Fig.$2$, when the activity of transactions takes place at the time step $i$.
We use two sets of tick data(KTB$203$ and KTB$209$) composed of bond futures prices,
and the average time between ticks for these is about one monute.

From now on, let's consider three return states composed by $N$ agents,
i.e., the continuous tick data of bond futures price.
We assume that the states of agent $l$ can be constituted into $\psi_l = \lbrace -1, 0, 1 \rbrace $,
where the state of clusters is given by $s(t_i ) =   \sum_{l=1}^{N} \psi_l $.
Thus the waiting state that occurs no transactions or gets no return corresponds to $\psi_l =0 $, 
and the selling and buying states, i.e., the active states of transaction, are $\psi_l =1$ and $\psi_l = -1$, 
respectively.
The active states of transaction are represented by vertices in a network having links of time series, and we assume that it belongs to
the same cluster between a group of agents sharing the same information and making a common decision. 
Since the distribution of returns is really related to the distributon of cluster, we can obtain the averaged distribution of
cluster in our model. Fig.$3$ presents the log-log plot of the averaged distribution of
cluster as a function of the size of the transacted states, and these scale with the scaling exponent $\alpha =1.75$. 

In order to find the distribution of the price return $R$ for different herding probabilities,
the network of links for $R$ can be selected at random value $a=a_{+} + a_{-} $,
where $a_{+}$ and $ a_{-}$ is, repectively, probability of the selling and buying herds.
As the herding parameter is defined by $ h \equiv \frac{1-a}{a}$,     
the probability distribution of returns for three types of herding parameter
scales as a power law $R^{-\beta}$ with the exponents $ \beta=3.6$(KTB$203$) and $2.9$(KTB$209$), 
as shown in Fig.$4$ (a) and (b).
Here we would suggest that $h^{*} =2.33$$(a=0.3)$ is the so-called critical herding parameter
from our data. It is obtained that the financial crashes occur at $a<0.3$($h>2.33$).
Thus the crash regime appears to increase in the probability with high returns values,
since the state of transaction exists to decrease lesser. 

Particularly, we can calculate the distribution of normalized returns for the bond futures price since the normalized return
is given by $(R-<R>)/\sigma$, where $<R>$ is the value of returns averaged over the time series
and the volatility $\sigma= (<R^2 > - <R>^2 )^{1/2}$.
In Fig.$5$ (a) and (b), we show, respectively, the semi-log plot of the probability distribution of the normalized returns 
in one case of herding probability $a=0.1$($h=9$), where the time steps $\Delta t=1$, $10$, $60$, and $360$
for two kinds of Korean treasury bond.  
Here the herd behavior for $\Delta t=1$, $10$, and $60$ is obtained to take the form of fat-tailed distribution of price returns.
On the other hand, the distribution of normalized returns really reduces to a Gaussian probability function
for the time step $\Delta t=360$. \\ 

\noindent
{\bf 3. CONCLUSIONS}
\hfill\\

In conclusion, we have investigated the dynamical herding behavior of the bond futures price
for two kinds of Korean treasury bond(KTB$203$ and KTB$209$) in Korean Futures Exchange market. 
Specially, the distribution of the price return for our bond futures scales as a power law
$R^{-\beta}$ with the exponents $ \beta =3.6$(KTB$203$) and $2.9$(KTB$209$). 
However, our distributions of the price return arenot in good agreement with those of Eguiluz and Zimmermann$^{5}$.
It is in practice found that our scaling exponents $ \beta $ are larger than theirs,
because our price returns are real values for KTB$203$ and KTB$209$.
It would be noted that the existence of financial crashes has extremely high probability 
for the active herding states making the same decision as the herding parameter takes the larger value.
We would also suggest that the critical value of herding parameter is $h^{*} =2.33$($a=0.3$),
and it shows a crossover toward a Gaussian probability function
for the distribution of normalized returns. 

In future, our results may be expected to be satisfactorily studied  
extensions of foreign financial analysis for the won-yen and won-dollar exchange rates
in Korean foreign exchange market.
We hope that it will apply our model of herd behavior to the other tick data 
in Korean financial markets and in detail compare our results with bond futures 
transacted in other nations. \\



%
%

%
%
\newpage
%
\begin{center}
{\bf FIGURE  CAPTIONS}
\end{center}

\vspace {10mm} 

\noindent
Figure $1$. Time series of bond futures price $P(t)$ for KTB$203$. 

\vspace {15mm}

\noindent
Figure $2$. Plot of the price return $R(t_i ) = $ ln$[P(t_i )/P(t_i -1)]$ for KTB$203$.

\vspace {15mm}

\noindent
Figure $3$.  Plot of the averaged probability distribution of cluster sizes $s$ 
for the herding probability
$a=0.5$ ($h=1$), where the averaged probability distributions for KTB$203$ and KTB$209$ scales as a power law 
$S^{-\alpha}$ with the exponent $ \alpha=1.75$(the dot line). 

\vspace {15mm}

\noindent
Figure $4$.  Log-log plot of the probability distribution of returns for three types of herding probabilities
$a=0.1$, $0.3$, $0.5$ ($h=9$, $2.33$, $1$), where the dot line scales as a power law 
$R^{-\beta}$ with the exponents (a) $ \beta=3.6$(KTB$203$) and (b) $2.9$(KTB$209$). 

\vspace {15mm}

\noindent
Figure $5$.  Semi-log plot of the probability distribution of the normalized returns 
($R/\sigma$) for herding probability $a=0.1$($h=9$), 
where the dot line is the form of Gaussian function with (a) $\sigma=1$ for KTB$203$ and
with (b) $\sigma=1$ for KTB$209$. 

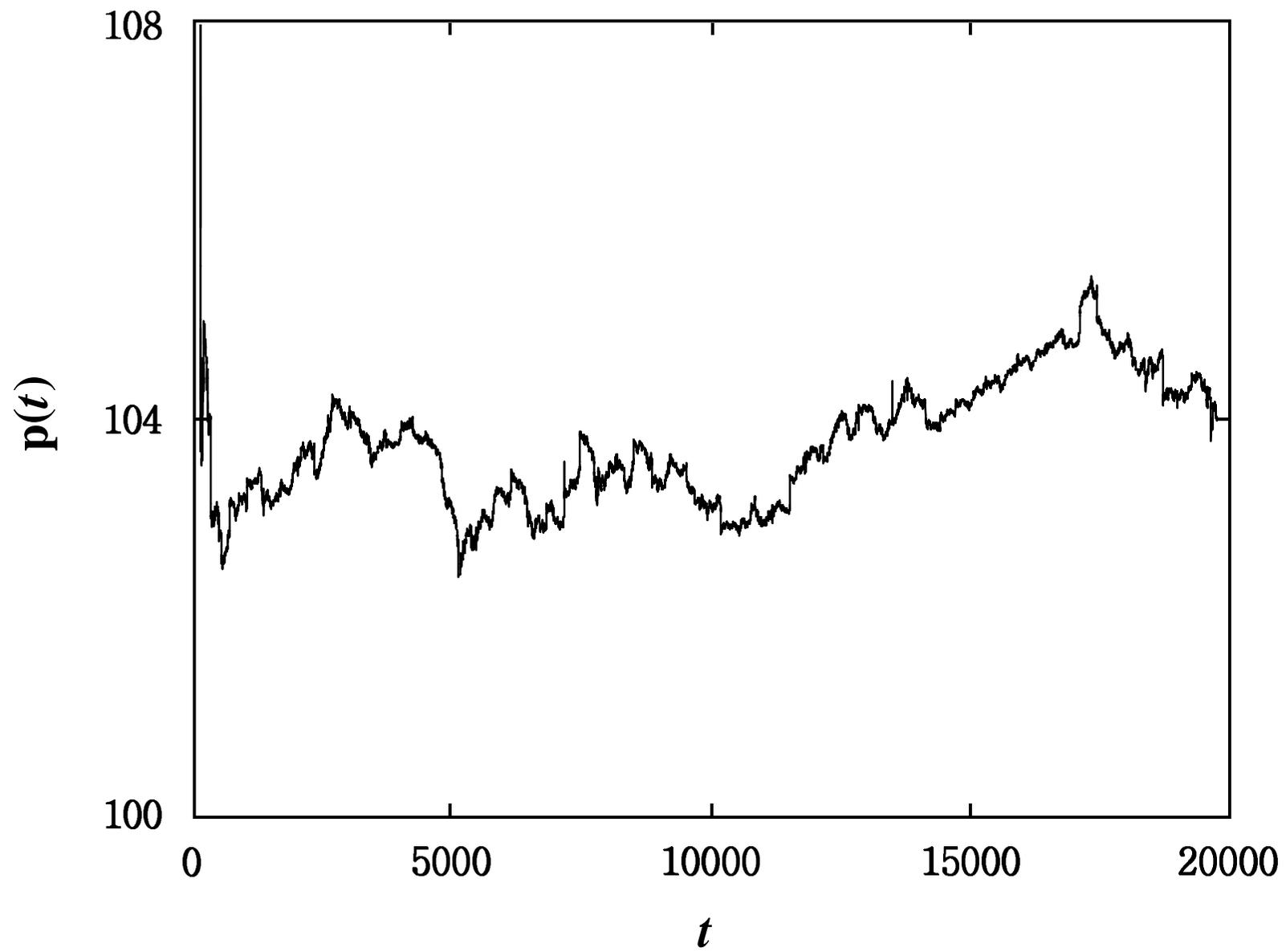
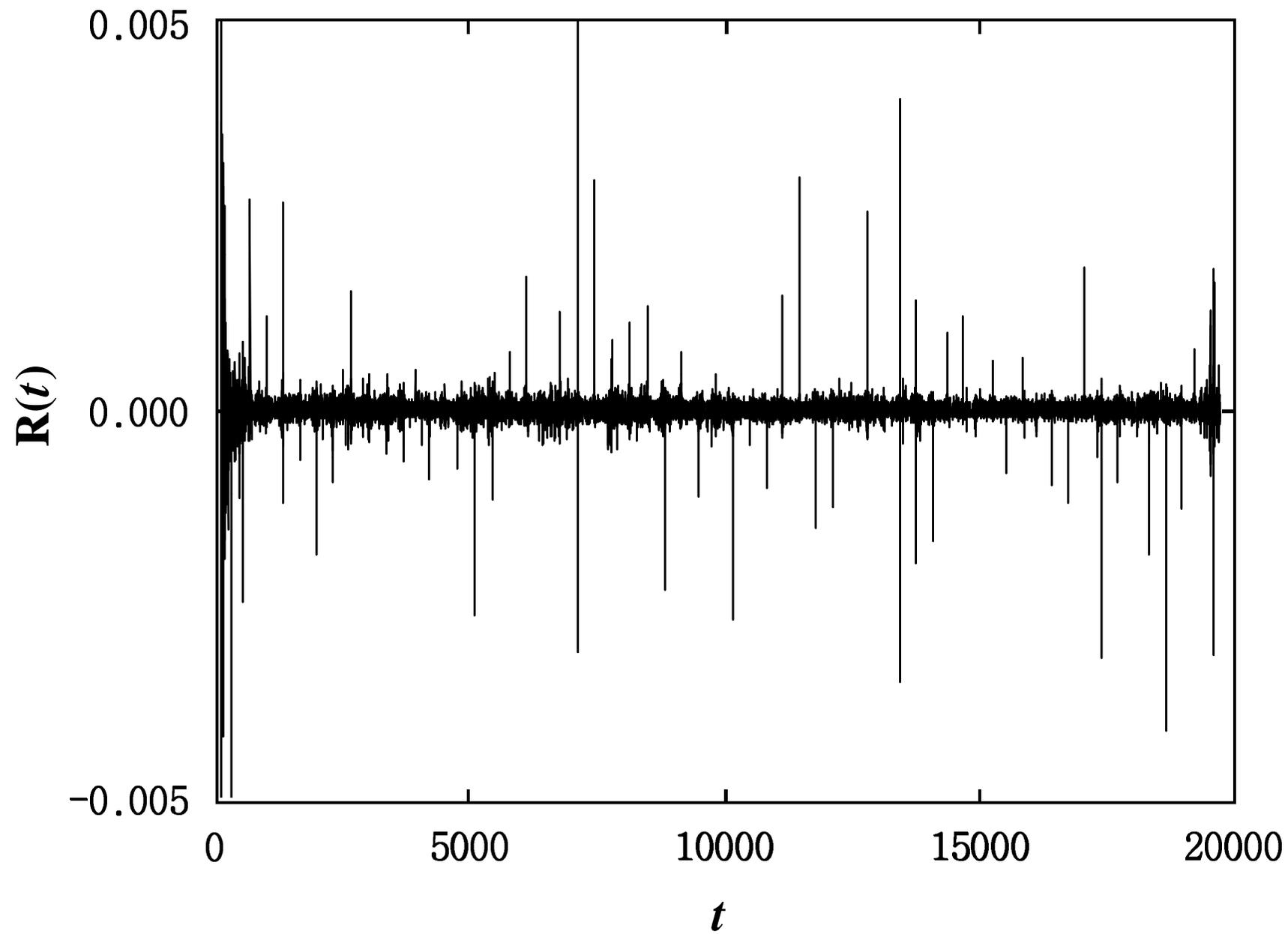
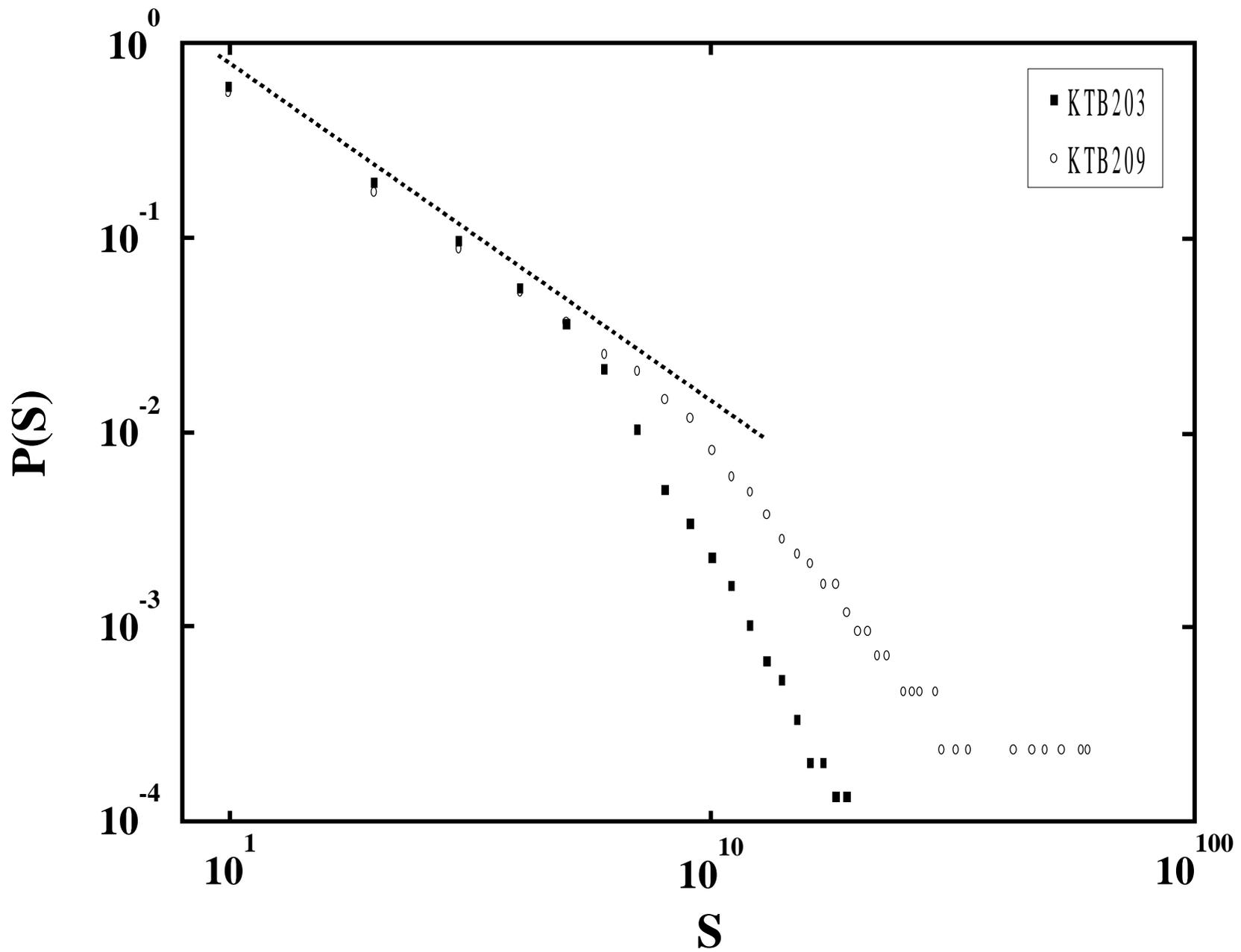
\end{document}